\documentclass{emulateapj}
\usepackage{multirow}
\usepackage[dvips]{color}  

\newcommand{\lya}{Lyman~$\alpha$}
\newcommand{\mgII}{Mg~II~2796, 2803}
\newcommand{\rmxaa}{Revista Mexicana de Astronomia y Astrofisica}
\newcommand{\cgsfnu}{erg~s$^{-1}$~cm$^{-2}$~Hz$^{-1}$}

\slugcomment{ApJ in press}

\shorttitle{Mg~II and Ly~$\alpha$ in galaxies}
\shortauthors{Rigby et al.}

\begin{document}

\title{On the lack of correlation between \mgII~\AA\ and Lyman~$\alpha$ emission in lensed star-forming galaxies}


\author{
J.~R.~Rigby\altaffilmark{1}, 
M.~B.~Bayliss\altaffilmark{2,3}
M.~D.~Gladders\altaffilmark{4,5}, 
K.~Sharon\altaffilmark{6}, 
E.~Wuyts\altaffilmark{7}, \&
H.~Dahle\altaffilmark{8}
}

\altaffiltext{1}{Astrophysics Science Division, 
           Goddard Space Flight Center, 8800 Greenbelt Rd., Greenbelt, MD 20771}
\altaffiltext{2}{Department of Physics, Harvard University, 
                 17 Oxford St., Cambridge, MA 02138}
\altaffiltext{3}{Harvard-Smithsonian Center for Astrophysics, 
                 60 Garden St., Cambridge, MA 02138}
\altaffiltext{4}{Department of Astronomy \& Astrophysics, University of
           Chicago, 5640 S. Ellis Ave., Chicago, IL 60637}
\altaffiltext{5}{Kavli Institute for Cosmological Physics, University of
          Chicago, 5640 South Ellis Ave., Chicago, IL 60637}
\altaffiltext{6}{Department of Astronomy, University of Michigan, 
          500 Church St., Ann Arbor, MI 48109}
\altaffiltext{7}{Max Plank Institute for Extraterrestrial Physics, 
                 Giessenbachstrasse 1, 85748 Garching, Germany}
\altaffiltext{8}{Institute of Theoretical Astrophysics, University of Oslo, 
              P.O. Box 1029, Blindern, NO-0315 Oslo, Norway}

\begin{abstract}
We examine the \mgII~\AA, Lyman $\alpha$, and nebular line emission
in five bright
star-forming galaxies at $1.66<z<1.91$ that have been gravitationally
lensed by foreground galaxy clusters.  All five galaxies show prominent
Mg~II emission and absorption in a P~Cygni profile.
We find no correlation between the equivalent widths of Mg~II and
\lya\ emission.  The Mg~II emission has a broader range of velocities 
than do the nebular emission line profiles; the Mg~II emission 
is redshifted with respect to systemic by
100 to 200 km~s$^{-1}$.  When present, \lya\ is even more redshifted.  The
reddest components of Mg~II and \lya\ emission have tails to 500--600~km~s$^{-1}$,
implying a strong outflow.  The lack of correlation in the Mg~II and \lya\
equivalent widths, the differing velocity profiles, and the high ratios of 
Mg~II to nebular line fluxes together suggest that the bulk of 
Mg~II emission does not ultimately arise as nebular line emission, but may instead be
reprocessed stellar continuum emission.
\end{abstract}


%
\section{Introduction}
It has recently become evident that the spectra
of distant galaxies frequently show \mgII\
in the classic P~Cygni profile of redshifted emission and
blueshifted absorption.  Such emission is notably absent in $z=0$ analogues.

\citet{Weiner09} reported the first detection of Mg~II emission in
distant galaxies, finding Mg~II emission with a P~Cygni profile
in the stacked spectra of z$=$1.4 galaxies.  They attributed this emission to a small
fraction (50 out of 1400) of the sample, ``most
probably from narrow-line active galactic nuclei'', in part because of
a high-velocity tail in the emission line profile.
\citet{Rubin10b} repeated this analysis on a
lower-redshift (0.7$<$z$<$1.5) sample of galaxies, and found Mg~II
emission in the stack of 468 galaxies, as well as in some 
individual galaxy spectra.  They suggested that this Mg~II
emission arises not from nuclear accretion, but from resonant scattering in an
expanding wind, analogous to the resonant scattering of
Lyman alpha.

\citet{Rubin11} found Mg~II emission that extended spatially 
across at least 7~kpc in a  z$=$0.69 galaxy, which also showed
prominent Fe~II* fine structure emission.  The authors noted that the
Fe~II* and Mg~II kinematics did not resemble those of nebular emission
lines.  
They argued that the Mg~II resonant
emission and the Fe~II* fine structure emission most likely arise via photon
scattering in an outflowing wind.
\citet{Giavalisco11} found Mg~II and Fe~II* emission in a stack of 
33 galaxies at z$=$1.6, and in a stack of 92 galaxies at 1.65$<$z$<$2.5.
They note that at matched spectral resolution, non-AGN z$=$0 galaxies 
show no such Mg~II and Fe~II* emission \citep{Kinney93}, except for
the Wolf-Rayet galaxy TOL 1924-416.
None of the galaxies in the \citet{Giavalisco11} stacks are 
detected in 4~Ms of Chandra X-ray integration, which argues against the
significant presence of X-ray--luminous AGN.

\citet{Erb12} showed how common
Mg~II and Fe~II* emission are in 1$<$z$<$2 galaxies.  Mg~II emission
is seen in one-third of their sample: 33 of 93 galaxies.  Fe~II* emission 
appears even more ubiquitous, appearing in all the stacks of subsamples.

\citet{Prochaska11} simulate Mg~II P Cygni profiles like those observed
in these distant galaxies,  using a Monte Carlo radiative transfer model 
that assumes a cool, outflowing wind.

Thus, while the first detections of Mg~II emission lines in
galaxies were thought to be a rare curiosity with a suspected AGN origin,
such emission is now recognized as common in distant star-forming
galaxies, though notably absent in z$=$0 analogues.  

In this paper, we examine Mg~II emission and absorption from five star-forming
galaxies at $1.66<z<1.91$
that have been gravitationally lensed by foreground galaxy clusters.  
While our sample size is much smaller than field samples, lensing magnification
has enabled us to obtain much higher--quality spectra of individual galaxies. 
We compare the Mg~II emission to the strengths and profiles of other emission
lines, to better understand where this Mg~II emission arises,
and what its physical origins are.

\section{Sample and observations}
The data analyzed in this paper
are part of a larger study of the rest-ultraviolet
spectral properties of bright lensed galaxies, which we are conducting with
the MagE spectrograph \citep{MagE} on the Clay Magellan II telescope.  
The galaxies discussed in this paper
are the five from our larger
sample that have redshifts such that both the \lya\ and \mgII\ features 
have spectral coverage.  
Four of the five galaxies discussed here are drawn from the SDSS Giant Arcs Sample
(SGAS),  a set of bright gravitationally lensed galaxies (Gladders
et al.\ in prep; \citealt{Bayliss11}, \citealt{Bayliss12}).  The other
galaxy, RCS0327 \citep{Wuyts10}, comes from the Second Red Sequence
Cluster Survey \citep{Gilbank11}.
Table~\ref{tab:obslog} gives total integration times and dates of the observations.
Further details of observations, data reduction, and full spectra will be
published in future papers.  

For RCS0327, we obtained MagE spectra for four distinct star-forming
knots, labeled E, U, B, G, following the nomenclature
of \citet{Sharon12}.  The spectrum for Knot E has the highest
signal-to-noise ratio.

\section{Results}

\subsection{Systemic redshift}\label{sec:systemic}
For three galaxies, we measured systemic redshifts 
by fitting the C III] 1907, 1909~\AA\
doublet with two Gaussians.  The fit was constrained such that both
lines shared a common linewidth.  The redshift was allowed to vary,
with the ratio of the central wavelengths held fixed.

S1441 lacks detected CIII] or other systemic lines; 
we therefore arbitrarily choose a systemic redshift of
z$=$1.666, which puts the reddest Mg~II 2796~\AA\ absorption at zero velocity.
RCS0327 has a complex velocity structure because it is undergoing a
major merger \citep{Wuyts14}.  
For Knot E of RCS0327 the C III] emission is blended with Fe II 2600 absorption
from an intervening system at z$=$0.98295;  we therefore
instead use Si~II*~1264, Si~II*~1309, Si~II*~1533, and Si~III]~1882, 
finding a weighted-average emission line redshift of 
$1.7034 \pm  0.00014$ for Knot E.
(This redshift is consistent within uncertainties with the section 
of the C III] profile that is not contaminated by intervening Fe II absorption.)
Table~\ref{tab:redshifts}  lists the measured systemic redshifts.

\subsection{Comparison of Ly$\alpha$ and Mg~II emission strength}

\citet{Erb12} reasoned  that Mg~II emission in z~$\sim$ 1--2 galaxies
should be similar to \lya\ emission, since both emission lines show P
Cygni profiles, and since their emission strengths are generally correlated
with UV color.  However, since they lacked spectral coverage of \lya, they could not
directly compare Mg~II and \lya\ 
emission profiles or equivalent widths.  

In Table~\ref{tab:EWs}  we report rest-frame equivalent widths and  2$\sigma$ upper limits 
for emission lines.  Reported uncertainties incorporate the
statistical uncertainty as well as the systematic uncertainty
due to continuum estimation. We determined the systematic uncertainty
empirically by comparing continuum fitting methods; different 
methods agree well and have a standard deviation that is typically less 
than half the measurement uncertainty. 

In Figure~1, 
we compare the equivalent widths of Mg~II
emission and \lya\ emission in our sample.  
The sample has reasonable dynamic range:
the equivalent widths of Mg~II vary by  a factor of 6, and
\lya\ by a factor of 8.
No correlation is observed.  

\subsection{Ubiquity of Mg II emission} 
While the five galaxies in this paper
were merely selected from our
larger sample as spectral coverage of Mg~II and \lya, in
fact, all five show detected Mg~II emission and absorption in a P
Cygni profile.  Mg~II spectra are plotted in
Figure~2. 
In \S\ref{sec:discussion} we explain why the ubiquity of Mg~II emission 
in our sample is likely a selection effect.

\subsection{Velocity structure of the emission}
We now examine the velocity structure of the Mg~II emission, compared
to other prominent spectral features.  Figure~2 
presents velocity plots for these five lensed galaxies.  
Though RCS0327 has spectra measured for four knots, in Figure~2 
we examine only the spectrum of Knot~E, since it has the highest
signal-to-noise ratio.  (Detailed analysis of knot-to-knot variations
in the Mg~II profile within RCS0327 is reserved for a future paper.)
Zero velocity
corresponds to the systemic redshift of each galaxy, as determined 
in \S\ref{sec:systemic}.

In each galaxy, the  Mg~II emission is broader than the C III]
and [O II] lines, and the peak is  redshifted with respect
to systemic by 100 to 200~km~s$^{-1}$. When Ly~$\alpha$ is detected, it is
even more redshifted than Mg~II, with peak intensity at 200--250~km~s$^{-1}$
from systemic.  

The red ``shoulder'' of Mg~II 2796~\AA\ emission will be absorbed by the
2803~\AA\ transition.  Thus, the 2803~\AA\ line is the one to examine
for redshifted emission.  The Mg~II profiles of S0108 and
RCS0327 Knot E show a ``shoulder'' of emission extending $\sim 500$~km~s$^{-1}$
redward of systemic.  The \lya\ emission in S0108 and S0957 also show
redshifted shoulders to even higher velocities ($>$600 ~km~s$^{-1}$).
Such extreme velocities imply a strong outflow.   
This result can be stated more generally:  all the strong, high
signal-to-noise Mg~II and \lya\ emission lines in our sample show 
high-velocity red ``shoulders''.

\section{Discussion and Conclusions}
\label{sec:discussion}
Mg~II emission is present in all five of the gravitationally--lensed
galaxies in our sample.  This may seem surprising,
given that \citet{Erb12} find such emission in only one-third of their
sample.   An explanation may lie in the fact that
Mg~II emission is stronger in the Erb et al.\ stacked subsamples with lower stellar
mass and bluer color; the reasons for this remain unclear.   
The lensed galaxies in our sample are quite blue.  
While we are still developing the lensing models required to infer intrinsic stellar
mass for this sample, galaxies selected in a similar way have 
stellar masses around $10^9$~M$_{\odot}$
\citep{Wuyts12},  
which would put them in the lower-mass bin of Erb et al.,
 the bin that showed strongest Mg~II emission.   Lensed arc
samples should be  biased toward galaxies with high surface brightness
features, in other words high specific star formation rates.  Given
the \citet{Erb12} results, it is not surprising that the
highly-magnified, high surface brightness star-forming regions
captured within the MagE slit should show Mg~II emission.

In RCS0327, the MagE spectra capture four different
star-forming knots, but in the other four galaxies, 
the MagE spectra capture only one bright knot of emission per galaxy.
Past work shows that such knots typically extend only a few 100 pc
\citep{Wuyts14,Sharon12, Jones12}.  
 In a future paper, we will examine the spatial variation of Mg~II 
emission within RCS0327.

The lack of a correlation between the equivalent widths of
Mg~II and \lya\ emission is striking, and can be interpreted in
several ways.
Since \lya\ is a bluer transition with a higher
absorption cross-section than Mg~II, it can be expected to be more heavily
obscured by dust.  Galaxy-to-galaxy variations in extinction
might erase an intrinsic correlation between Mg~II and \lya\
emission, if present. A similar decoupling between 
UV continuum and \lya\ emission was proposed by \citep{Giavalisco96}.
An alternate explanation for the lack of correlation would be that 
\lya\ and Mg~II emission
do not share common origins.  As a hydrogen recombination line, \lya\
originates in H~II regions, and is then resonantly scattered.   
The origin of Mg~II emission in distant galaxies is not well
established; the two models proposed in the literature are
intrinsic production of line emission in nebular regions with subsequent scattering, 
and intrinsic production of continuum photons that are resonantly scattered.
Plasma models of H~II regions (Cloudy, \citealt{Ferland13}) 
predict weak nebular Mg~II emission
(\citealt{Erb12}, their Figure 16); the dominant emission mechanism
is the excitation of Mg$^+$ by electron impact (G. Ferland, priv. comm.)   
Thus, the flux of any \textit{nebular} component of 
Mg~II emission should be tied to the electron temperature and luminosity 
of the H~II  regions and the \lya\ flux, and
 not tied to the amount of Mg~II absorption.  
A future study could test for a correlation between the extinction-corrected 
H $\alpha$ luminosity, which traces star formation rate, and the 
luminosity of Mg~II emission.

Nebular Mg~II emission is observed in the local universe, 
in the Orion H~II
region complex, as reviewed by  \citet{Dufour87}.\footnote{Several
authors \citep{Rubin11,Prochaska11,Erb12} have erroneously
  cited \citet{Kinney93} as claiming that Mg~II can be produced in H~II
  regions.  Kinney et al.\ in fact claimed no such detection, but rather
  adopted a list of emission lines seen in H~II regions by IUE, as
  reviewed by \citet{Dufour87}.  The correct citations should be 
\citet{Torres-Peimbert80} and \citet{Boeshaar82}, as reviewed by \citet{Dufour87}.  
We note this to correct the citation   stream.}
\citet{Torres-Peimbert80}  detected Mg~II emission at
 low resolution with the International Ultraviolet Explorer (IUE) 
for a part of Orion; the flux of the emission
 was comparable to but less than C~III] 1909.
In high-resolution spectra of six regions  of Orion, \citet{Boeshaar82} 
find that the Mg~II emission ranges from 
 0.04 to 1.35 times the strength of [O II] 2470~\AA, and 
 0.25 to 25 times the strength of the C III] 1907,1909~\AA\ doublet.
By comparison, in our five galaxies, Mg~II emission is much stronger 
than both [O II]~2470 and C~III]~1907. 
These high line flux ratios argue against an intrinsically 
nebular origin for the  bulk of the Mg~II emission.

Mg~II emission could also arise through the scattering of continuum photons 
that have the wavelength of the Mg~II doublet, and are resonantly scattered
through a velocity gradient until they emerge as Mg~II emission. 
If the ultimate origin of Mg~II emission is from continuum photons, then the 
equivalent width of Mg~II emission should have no
connection to that of \lya, but instead be tied to the Mg~II absorbing
column density and the characteristics of the outflowing wind.
This could be examined in future work, for example by testing for a correlation between 
Mg~II emission and the intrinsic \lya\ emission inferred from the 
 observed fluxes of Lyman alpha and two Balmer lines.

The galaxies in our sample show a complex velocity structure in the emission
components, with Mg~II redshifted with respect to the nebular and
Fe~II* lines, and \lya\ (when present) even more redshifted, with a
tail to more than 600~km~s$^{-1}$ from systemic.  These observations
can provide new insights into the wind structure of z$\sim$2 star-forming
galaxies, through future multi-ion wind modeling.  
For now, it is enough to say that the Mg~II emission
does not have the velocity structure of either the nebular emission or the
\lya\ emission.  This suggests that the bulk of Mg~II emission 
is emitted from physically different regions of the galaxy than is the nebular or \lya\ 
emission.

To conclude, the rest-ultraviolet spectra of five lensed galaxies presented here
reveal that Mg~II emission is not a simple proxy for Lyman $\alpha$ emission;
its equivalent width is uncorrelated with that of \lya, and its velocity structure
is different as well.  Mg~II emission, since it is  bright and common in $z\sim2$ galaxies, 
may prove to be a  powerful diagnostic of winds driven by star formation
in distant galaxies, once more work is done to understand its physical origin and
radiative transfer.

\acknowledgments
Acknowledgments:  This paper includes data gathered with the 6.5
meter Magellan Telescopes located at Las Campanas Observatory, Chile.  
Magellan time for this project was granted by the Carnegie
Observatories, U.~Michigan, U.~Chicago, and the Harvard-Smithsonian
Center for Astrophysics.

\begin{deluxetable*}{lll}
\tablecolumns{3}
\tablewidth{0pc}
\tablenum{1}
\label{tab:obslog}
\tablecaption{Log of Observations}
\tablehead{
\colhead{source} & \colhead{t (hr)} & \colhead{UT date}
}
\startdata
SGAS~J000451.7$-$010321    &  8.38   & 2010-12-09, 2013-08-10,  2013-10-06, 2013-10-07\\
SGAS~J010842.2$+$062444    &  7.50  & 2012-08-19, 2012-08-20, 2012-09-14, 2013-08-10\\
RCSGA 032727$-$132609 knot E & 10.0   & 2008-07-31, 2010-12-09, 2010-12-10\\
 ~~~knot U                       & 7.94    & 2008-07-31 , 2010-02-16, 2010-02-17, 2013-10-06, 2013-10-07\\
 ~~~knot B                        & 2.83   &  2013-10-06,  2013-10-07\\
 ~~~knot G                        & 1.83   & 2013-10-06, 2013-10-07\\
SGAS~J095738.7$+$050929 &  2.08   & 2010-12-09, 2010-12-10\\ 
SGAS~J144133.2$-$005401  &   1.50  & 2009-04-22\\
\enddata
\tablecomments{Source names, total integration times, and UT dates of observation.  
Full source names follow the convention of
SGAS~JHHMMSS.s$\pm$DDMMSS \citep{Koester10};
abbreviated names (HHMM) are used subsequently}.
\end{deluxetable*}

\begin{deluxetable*}{lll}
\tablecolumns{3}
\tablewidth{0pc}
\tablenum{2}
\label{tab:redshifts}
\tablecaption{Systemic Redshifts}
\tablehead{
\colhead{source} & \colhead{systemic redshift z$_s$} & \colhead{emission
  lines used for z$_s$}}
\startdata
S0004  &  $1.6811   \pm   0.0001$  &  C III] 1907/1909\\
S0108  &  $1.91021 \pm   0.00003$  &  C III] 1907/1909\\
R0327 knotE & $1.7034 \pm  0.00014$ &  Si~II*~1264, 1309, 1533,  Si~III]~1882\\
S0957  &  $1.82042  \pm 0.00004$   & C III] 1907/1909\\
\enddata
\end{deluxetable*}

\begin{deluxetable}{lllllllll}
\tablecolumns{9}
\tablewidth{0pc}
\tablenum{3}
\label{tab:EWs}
\tablecaption{Measured equivalent width of emission lines}
\tablehead{\colhead{source} & \colhead{2796} &  \colhead{2803} &  \colhead{Ly $\alpha$} & 
\colhead{1907} & \colhead{1909} & \colhead{2326} & \colhead{2396} & \colhead{2470}} 
\startdata 
 S0004          & $-0.45 \pm 0.1$ & $-0.81 \pm 0.17$ & $-3.3 \pm 0.4$ & $-0.2\pm0.1$  &  $-0.25\pm0.1$ &  $>-0.26$      & $>-0.26$         &  $>0.26$  \\
S0108          & $-0.9  \pm 0.6$  & $-0.87 \pm 0.25$ & $-8.9 \pm 0.6$ & $-1.2\pm0.1$  &  $-0.7\pm0.1$  &  $-0.5\pm0.1$  &  $-0.55\pm0.15$  &  $>-0.30$ \\                  
R0327 knot E & $-1.3  \pm 0.16$ & $-1.00 \pm 0.17$ & $>-1.2$          & $-0.8\pm0.1$  &  $-1.2\pm0.1$  &  $-0.6\pm0.15$ &  $-0.8 \pm0.15$  &  $-1.0\pm0.1$ \\            
~~~knotB     & $-1.1 \pm 0.5$   &  $-0.8 \pm 0.3$    &  $>-2.5$       & $-1.6\pm0.4$  &  $-1.4\pm0.35$ &  $>-0.7$       &  $>-0.8$         &  $-1.1\pm0.35$ \\            
~~~knotG    & $-1.6 \pm 0.6$   &  $-1.4 \pm  0.6$    &  $>-3.8$       & $>-1.1$       &  $>-1.05$      &  $>-0.96$      &  $-0.8\pm0.4$    &   $>-0.75$ \\                
~~~knotU    & $-2.8 \pm 0.3$   &  $-2.35 \pm 0.2$    &  $>-1.1$       & $-1.5\pm0.1$  &  $-0.9\pm0.1$  &  $-1.1\pm0.4$  &  $-1.1\pm0.2$    &   $-0.9\pm0.1$\\                
S0957          & $-2.8  \pm 0.9$  & $-1.81 \pm 0.77$ & $-8.1 \pm 1.2$ & $-2.0\pm0.3$  &  $-1.2\pm0.3$  &  $>-0.87$      &  $>-0.71$        &   $>-0.85$  \\ 
S1441          & $-0.95 \pm 0.4$  & $>-0.8$          & $>-3.4$        & $>-0.62$      &  $>-0.62$      &  $>-1.0$       &  $>-1.2$         &  $>-0.92$  \\
\enddata
\tablecomments{Measured rest-frame equivalent width, in \AA, of the following emission lines:
Mg~II 2796, Mg~II~2803, Lyman $\alpha$, C~III] 1907,  C~III] 1909, C~II]~2326, Fe~II*~2396, [O II]~2470.
For lines with P~Cygni profiles, equivalent widths are of the emission component only. Two $\sigma$ upper limits are quoted for non-detections.}
\end{deluxetable}

\begin{figure}
\includegraphics[width=5in,angle=-90]{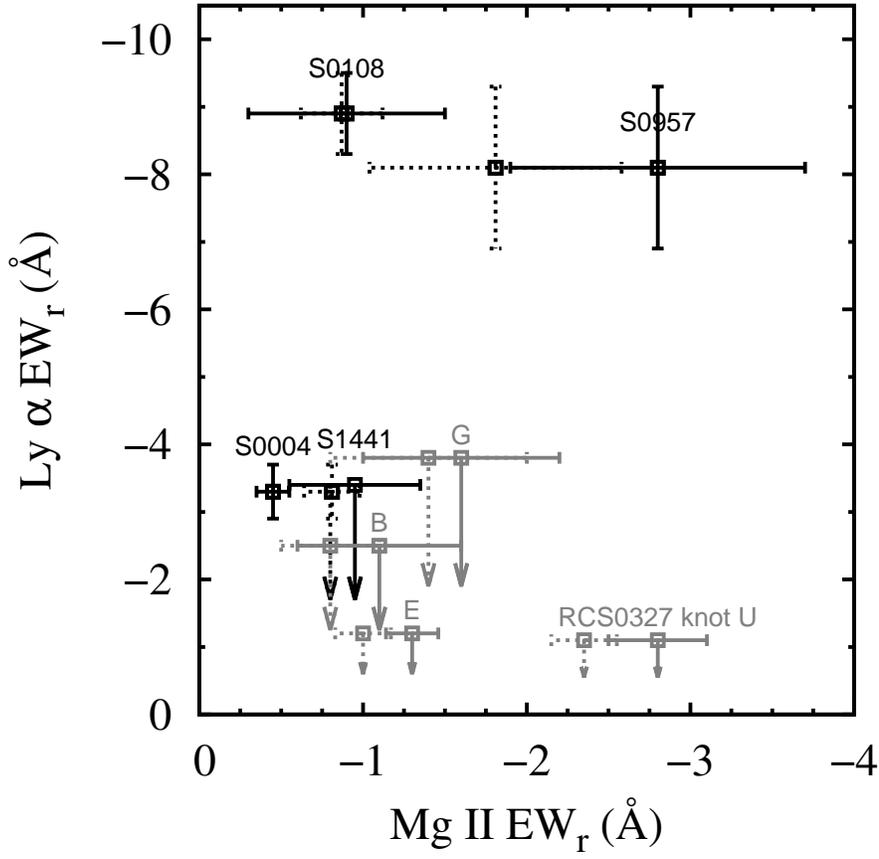}
\figcaption{Rest-frame equivalenth width of Lyman $\alpha$ emission,
  as a function of the rest-frame equivalent widths of Mg~II 2796~\AA\  emission 
\textit{(errorbars plotted as solid lines)}, 
and Mg~II 2803~\AA\ emission \textit{(errorbars plotted with dotted lines)}.  
Values for each of four star-forming knots of RCS0327
are plotted in grey.   The other galaxies are plotted in black.
There is no apparent correlation between the equivalent widths of Mg~II and
  Ly$\alpha$.
}
\label{fig:LyaMgEW}
\end{figure}

\begin{figure*}
\includegraphics[width=5.5in,angle=-90]{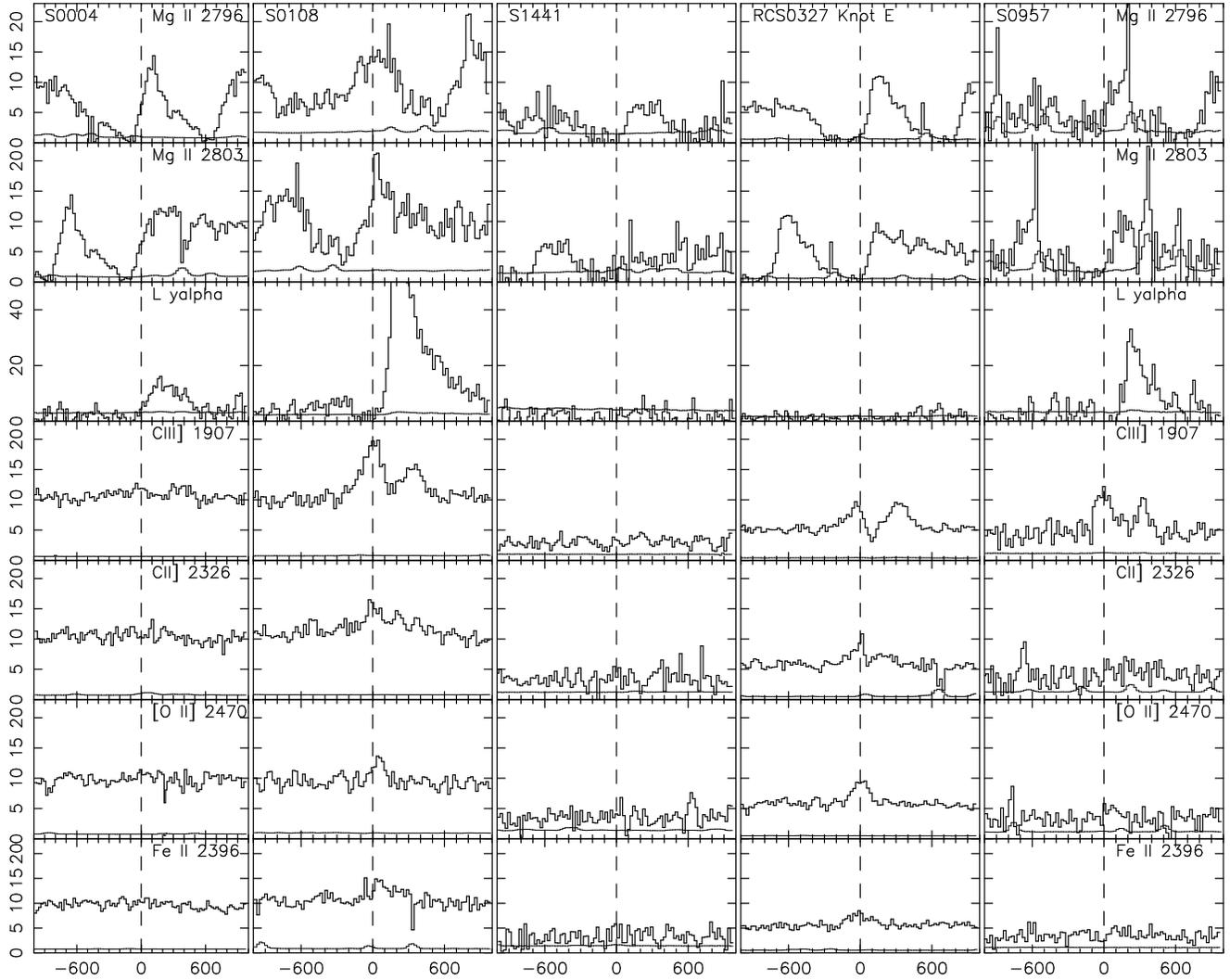}
\figcaption{Velocity plots of spectral features in the MagE
 spectra of five lensed galaxies.
Each column plots the spectrum of a different galaxy, labeled at top, as well
as its $1\sigma$ uncertainty spectrum. 
The x-axis is velocity in km/s; zero velocity should be the systemic
redshift.  The y-axis is flux density in $1\times 10^{-29}$~\cgsfnu.    
Each row plots a different emission line, labeled at left and right.
For RCS0327, only the spectrum of Knot E is shown.  
}\label{fig:spectra}
\end{figure*}

\end{document}